\documentstyle[preprint,aps]{revtex}
\begin{document}
\draft
\title{Magnetic Properties of a Quantum Ferrimagnet: NiCu(pba)(D$_{2}$O)$_{3}$$\cdot$2D$_{2}$O}
\author
{ 
Masayuki {\sc Hagiwara}$^{1,2}$\footnote{e-mail: 
hagiwara@postman.riken.go.jp}, Kazuhiko {\sc Minami}$^{3}$, Yasuo {\sc 
Narumi}$^{4,2}$, Keiji {\sc Tatani}$^{2}$ \\ and Koichi {\sc Kindo}$^{2,4}$}
\address{$^1$The Institute of Physical and Chemical Research (RIKEN),
Wako, Saitama 351-0198 \\ 
$^2$KYOKUGEN, Osaka University, Toyonaka 560-8531 \\ 
$^3$Graduate School of Mathematics, Nagoya University, Nagoya 464-8602 \\ 
$^4$CREST, Japan Science and Technology Corporation (JST), Kawaguchi, 
Saitama 332-0012 \\}
\date
{Received May 1, 1998}
\maketitle
\begin{abstract}
We report the results of magnetic measurements on a powder sample of 
NiCu(pba)(D$_{2}$O)$_{3}$$\cdot$2D$_{2}$O (pba=1,3-propylenebis(oxamato)) 
which is one of the prototypical examples of an $S$=1/2 and 1 ferrimagnetic chain.   
Susceptibility($\chi$) shows a monotonous increase with decreasing 
temperature (T) and reaches a maximum at about 7 K.    In the plot of 
$\chi$$T$ versus $\it{T}$, the experimental data exhibit a broad 
minimum and are fit to the $\chi$$T$ curve calculated for the ferrimagnetic Heisenberg 
chain composed of $S$=1/2 and 1.   From this fit, we have evaluated the 
nearest-neighbor exchange constant $J/k_{\rm B}$=121 K, the g-values of 
Ni$^{2+}$ and Cu$^{2+}$, $g_{\rm{Ni}}$=2.22 and $g_{\rm{Cu}}$=2.09, respectively.
Applied external field dependence of $\chi$$T$ at low temperatures is 
reproduced fairly well by the calculation for the same ferrimagnetic model.        
\end{abstract}

Extensive studies of one-dimensional systems were prompted by Haldane's 
theoretical work~\cite{haldane} in 1983 after the initial wave of 
studies~\cite{jongh,steiner} in the late 
1960s and early 1970s.   Recently, quantum spin systems with singlet 
ground states, namely Haldane systems~\cite{haldane} (linear chain Heisenberg 
antiferromagnets with integer spin values), inorganic spin-Peierls systems~\cite{hase} 
and even-leg spin ladder systems,~\cite{dagotto} have been studied 
extensively.   In particular, 
cuprate systems have attracted much attention because of the relation to 
high $\it{T}_{\rm{c}}$ superconductors.   

In regard to the one-dimensional systems with magnetic ground states, an $S$=1/2 and 1 ferrimagnetic chain has been theoretically investigated 
recently,~\cite{kolezhuk,pati,brehmer,kuramoto} in addition to some pioneering theoretical 
works~\cite{drillon,drillon2} published in the 1980s.  From the low dimensionality and small spin values in this 
system, we expect a kind of quantum effect.   Theoretical 
studies show some remarkable features as follows: (1) between two 
low-lying, gapless and gapped excitation branches, the gapped branch lies higher than that deduced from 
a conventional spin wave theory.   From reliable calculations,~\cite{brehmer} the 
gap ($\Delta$/$J$) has been evaluated to be 1.767$\pm$0.003 where $J$ is 
the nearest-neighbor exchange constant.  The definition of the Hamiltonian 
will be shown later.   (2) The spin correlation length between 
sublattice moments is extremely short.  The length is below unit cell 
length and can not be evaluated with numerical accuracy.   (3) The full 
magnetization curve up to saturated magnetization is calculated and 
is obviously different from that for a classical ferrimagnet.~\cite{matsuura}    

On the other hand, although some candidates for the ferrimagnetic 
Heisenberg chain composed of spin 1/2 and 1 exist in real 
bimetallic substances,~\cite{pei,koningsbruggen} only preliminary 
magnetic measurements and comparisons with numerical calculations were 
made.~\cite{pei}   Thus, we investigate precisely the magnetic properties of an alternating Ni and Cu chain compound 
NiCu(pba)(D$_{2}$O)$_{3}$$\cdot$2D$_{2}$O 
(pba=1,3-propylenebis(oxamato)).  In these measurements, we use a deutrated sample 
because it is of superior quality to a hydrated one, although 
reason for this remain unclear and we plan to perform neutron scattering 
measurements on this deutrated sample.   The format used in this letter is 
as follows: in the next section, we discuss the synthesis and crystal 
structure of NiCu(pba)(D$_{2}$O)$_{3}$$\cdot$2D$_{2}$O.  We then report 
the results of magnetic measurements and of the comparison with 
numerical calculations for $S$=1/2 and 1 ferrimagnetic Heisenberg chain.   
Finally, we show the field dependence of $\chi$ times $\it{T}$ and compare the 
experimental data with calculated ones for the same model.      

Powder samples of NiCu(pba)(D$_{2}$O)$_{3}$$\cdot$2D$_{2}$O were synthesized according to 
the procedure reported in ref.~13.   Na$_{2}$[Cu(pba)]$\cdot$6H$_{2}$O was prepared from 
CuSO$_{4}$, NaOH and 1,3-trimethylenebis(oxamido)~\cite{nonoyama} which was previously 
synthesized from ethyl oxamate and 1,3-propanediamine.   Then, the title 
compound was obtained by slow diffusion of aqueous solutions (D$_{2}$O 
$>$99.8\%) of Na$_{2}$[Cu(pba)]$\cdot$6H$_{2}$O and 
Ni(ClO$_{4}$)$_{2}$$\cdot$6H$_{2}$O in a U-tube.  Chemical analysis showed 
a slight deviation of H content from the ratio in the ideal deutrated 
sample, but the molecular weight of this sample was only about 1\% smaller than 
that of NiCu(pba)(D$_{2}$O)$_{3}$$\cdot$2D$_{2}$O.    
Crystal structure of NiCu(pba)(H$_{2}$O)$_{3}$$\cdot$2H$_{2}$O has not been 
analyzed, but that of a similar compound 
MnCu(pba)(H$_{2}$O)$_{3}$$\cdot$2H$_{2}$O where Mn replaces Ni, has been 
analyzed.~\cite{pei}  Powder x-ray
diffraction patterns of these compounds show that these belong to the same 
space group.   Thus, NiCu(pba)(H$_{2}$O)$_{3}$$\cdot$2H$_{2}$O crystallizes in the orthorhombic system and belongs to 
the $Pnma$ space group.~\cite{pei}   As shown in Fig.~1, the structure consists of ordered 
bimetallic chains along the $\it{b}$ axis with octahedral Ni$^{2+}$ and 
square-pyramidal Cu$^{2+}$ ions bridged by oxamato groups.   
At the apical positions of Ni and Cu, water molecules are bound.              

Magnetic measurements were carried out with a SQUID 
magnetometer (Quantum Design's MPMS-XL7L) at KYOKUGEN in Osaka University.   
We show in Fig.~2 the dc magnetic susceptibility $\chi$(=$M$/$H$ where 
$M$ and $H$ 
represent magnetization of the sample and the external magnetic field, 
respectively) of a powder sample of NiCu(pba)(D$_{2}$O)$_{3}$$\cdot$2D$_{2}$O.   
The susceptibility of NiCu(pba)(D$_{2}$O)$_{3}$$\cdot$2D$_{2}$O 
increases monotonously with decreasing temperature until about 7 K, at which the 
susceptibility reaches a maximum.  Below 7 K, the long-range order 
probably occurs due to the interchain couplings.     

Figure 3 shows $\chi$ times $\it{T}$ of NiCu(pba)(D$_{2}$O)$_{3}$$\cdot$2D$_{2}$O as a function of temperature. 
This plot is familiar to chemists but not to physicists.   Therefore, 
we explain this plot in some details.   If a magnetic system is paramagnetic, 
$\chi$$T$ is constant over the whole temperature range.  If a magnetic 
system has a dominant ferromagnetic (antiferromagnetic) interaction, 
$\chi$$T$ increases (decreases) when the temperature is decreased.   In Fig.~3, 
$\chi$$T$ decreases when decreasing the temperature from 300 K, implying that antiferromagnetic 
coupling exists between the nearest neighbor spins, and reaches a rounded minimum 
at about 70 K.   Then, $\chi$$T$ increases and reaches a maximum at about 10 
K, and hereafter, when the temperature is decreased further, it decreases rapidly.   
The increase in $\chi$$T$ below 70 K implies that this ferrimagnetic system 
behaves like a ferromagnetic chain at low temperatures.  
Interchain (antiferromagnetic) couplings probably  give rise to the steep 
decrease of $\chi$$T$ below 10 K.   We compare the experimental data 
with numerical calculations (exact diagonalization method up to five unit 
cells (ten sites)) for the $S$=1/2 and 1 
ferrimagnetic Heisenberg chain.   The 
Hamiltonian of this system in a magnetic field is defined by
\begin{equation}
   {\cal H} = J \sum_{i=1}^{L}{[{\bf S_{i}} \cdot {\bf s_{i}}+{\bf s_{i}} 
   \cdot {\bf S_{i+1}}]}\\
       -g_{S} \mu_{\rm B}H \sum_{i=1}^{L} {{\bf S_{i}}}-g_{s} \mu_{\rm 
       B}H \sum_{i=1}^{L} {{\bf s_{i}}}, 
\end{equation}
where {\bf S} and {\bf s} are the $S$=1 and $S$=1/2 spin operators, 
respectively, and $g_{\rm S}$ and 
$g_{\rm s}$ the g-values of the $S$=1 and $S$=1/2 magnetic 
moments, respectively, and $\mu_{\rm B}$ the Bohr magneton and $\it{H}$ the 
external magnetic field.   Here, the periodic boundary condition is 
imposed, so that {\bf S}$_1$={\bf S}$_{L+1}$.   The solid line in Fig.~3 
shows the result of the best fit to the experimental data between 30 K 
and 150 K.  Good agreement between experimental and calculated results 
is achieved between 30 K and 150 K.   Slight deviation at high 
temperatures may arise 
from the error enhancement of $\chi$$T$ at high temperatures or 
the omission of the single ion anisotropy term which exists in Ni($S$=1) 
compounds.   From this fit, we obtain the exchange constant $J/k_{\rm B}$=121 
K, $g_{\rm S}$(=$g_{\rm{Ni}}$)=2.22 and $g_{\rm s}$(=$g_{\rm{Cu}}$)=2.09.    

Next, we show the magnetic field dependence of $\chi$$T$ in Fig.~4.
Experimental data of $\chi$$T$ at 0.1 T (open squares), 1 T (open 
triangles) and 7 T (open circles) 
are plotted in the upper panel. Experimental data at 0.1 T and 1 T have a 
similar tendency, but $\chi$$T$ at 7 T at low temperatures deviates 
significantly from the others.   This behavior is reproduced in the calculation shown 
in the lower panel.   Here, the designated $H$/$J$ figures represent 
those of $g\mu_{\rm B}H$/$J$ 
with $g$=2.0 and $J/k_{\rm B}$=121 K, and magnitudes of $\chi$$T$ are 
calculated for the above Hamiltonian using $g_{\rm S}$=2.22 and 
$g_{\rm s}$=2.09.  The low temperature behavior of $\chi$$T$ at 7 T implies that 
the N$\acute{e}$el order tends to be fixed 
and the ferromagnetic fluctuation of magnetic moments is suppressed by the external field.     

In conclusion, magnetic properties of alternating Ni and Cu chain compound 
NiCu(pba)(D$_{2}$O)$_{3}$$\cdot$2D$_{2}$O were investigated by 
magnetic susceptibility measurements.   From comparison with 
 a numerical calculation for the ferrimagnetic Heisenberg chain composed 
 of $S$=1/2 and 1, we have obtained the values 
of the exchange constant and the $g$-values of Ni and Cu.    Field 
dependence of $\chi$$T$ at low temperatures has been reproduced by similar 
calculations in magnetic fields.   

\section*{Acknowledgments}
This work was carried out under the Visiting Researcher$^{\prime}$s Program of 
KYOKUGEN at Osaka University and was supported in part by a 
Grand-in-Aid for Scientific Research from the Ministry of Education, 
Science, Sports and Culture.  M.H. would like to thank Professor K. 
Nonoyama of Konan Women$^{\prime}$s Junior College for information about the synthesis of 
Na$_{2}$[Cu(pba)]$\cdot$6H$_{2}$O and Professor S. Yamamoto of Okayama 
University for fruitful discussions.  Thanks are also due to the Chemical Analysis
Units in RIKEN.

\begin{figure}
 \caption{Crystal structure of 
 NiCu(pba)(H$_{2}$O)$_{3}$$\cdot$2H$_{2}$O.   Alternating Ni and Cu chain 
 runs along the $\it{b}$ axis.   Hydrogen atoms are omitted for clarity.}
\label{fig1}
\end{figure}

\begin{figure}
 \caption{Temperature dependence of the susceptibility(=$M$/$H$) of a powder 
 sample of NiCu(pba)(D$_{2}$O)$_{3}$$\cdot$2D$_{2}$O.  Inset: Susceptibility of 
 NiCu(pba)(D$_{2}$O)$_{3}$$\cdot$2D$_{2}$O at low temperatures.  We see 
 the cusp around 7 K where the long-range order probably occurs.}
\label{fig2}
\end{figure}

\begin{figure}
 \caption{$\chi$$T$ versus temperature plot of NiCu(pba)(D$_{2}$O)$_{3}$$\cdot$2D$_{2}$O.   
 The solid line shows the fit to a numerical calculation for the 
 ferrimagnetic Heisenberg chain composed of $S$=1/2 and 1.}  
 \label{fig3}
 \end{figure}
 
 \begin{figure}
 \caption{Magnetic field dependence of $\chi$$T$ of 
 NiCu(pba)(D$_{2}$O)$_{3}$$\cdot$2D$_{2}$O.  Upper panel shows the 
 experimental data for the designated applied field.  The values of $H$/$J$ 
 represent the ratios of $g\mu_{\rm B}H$ to $J$ with $g$=2.0.  In the lower 
 panel, field dependence of calculated $\chi$$T$ corresponding to $H$/$J$ 
 values in the upper panel is shown. }  
 \label{fig4}
 \end{figure}   
 
\end{document}